\documentclass[prc,preprint,showpacs,preprintnumbers,amsmath,amssymb]{revtex4}

\usepackage[mathscr]{eucal}
\usepackage{graphicx}
\def\slr#1{\setbox0=\hbox{$#1$}           
   \dimen0=\wd0                                 
   \setbox1=\hbox{/} \dimen1=\wd1               
   \ifdim\dimen0>\dimen1                        
      \rlap{\hbox to \dimen0{\hfil/\hfil}}      
      #1                                        
   \else                                        
      \rlap{\hbox to \dimen1{\hfil$#1$\hfil}}   
      /                                         
   \fi}

\def\ksq{k^2}

\def\mytint#1{\!\int\!\!\frac{d^3\!{#1}}{(2\pi)^3}\,}
\def\gev#1{ GeV${}^{#1}$}
\def\be{\begin{eqnarray}}
\def\ee{\end{eqnarray}}

\renewcommand{\theequation}%
    {\arabic{section}.\arabic{equation}}
\makeatletter \@addtoreset{equation}{section} \makeatother

\begin{document}

\preprint{BCCNT: 03/101/319}

\title{Comparison of Temperature-Dependent Hadronic Current Correlation
Functions Calculated in Lattice Simulations of QCD and with a
Chiral Lagrangian Model}

\author{Bing He}
\author{Hu Li}
\author{C. M. Shakin}
 \email[email:]{casbc@cunyvm.cuny.edu}
\author{Qing Sun}

\affiliation{%
Department of Physics and Center for Nuclear Theory\\
Brooklyn College of the City University of New York\\
Brooklyn, New York 11210
}%

\date{February, 2003}

\begin{abstract}

The Euclidean-time hadronic current correlation functions,
$G_P(\tau, T)$ and $G_V(\tau, T)$, of pseudoscalar and vector
currents have recently been calculated in lattice simulations of
QCD and have been used to obtain the corresponding spectral
functions. We have used the Nambu-Jona-Lasinio (NJL) model to
calculate such spectral functions, as well as the Euclidean-time
correlators, and have made a comparison to the lattice results for
the correlators. We find evidence for the type of temperature
dependence of the NJL coupling parameters that we have used in
previous studies of the mesonic confinement-deconfinement
transition. We also see that the spectral functions obtained when
using the maximum-entropy-method (MEM) and the lattice data differ
from the spectral functions that we calculate in our chiral model.
However, our results for the Euclidean-time correlators are in
general agreement with the lattice results, with better agreement
when our temperature-dependent coupling parameters are used than
when temperature-independent parameters are used for the NJL
model. We also discuss some additional evidence for the utility of
temperature-dependent coupling parameters for the NJL model. For
example, if the constituent quark mass at $T=0$ is $352
\,\mbox{MeV}$ in the chiral limit, the transition temperature is
$T_c=208 \,\mbox{MeV}$ for the NJL model with a standard momentum
cutoff parameter. (If a Gaussian momentum cutoff is used, we find
$T_c=225 \,\mbox{MeV}$ in the chiral limit, with $m=368
\,\mbox{MeV}$ at $T=0$.) The introduction of a weak temperature
dependence for the coupling constant will move the value of $T_c$
into the range 150-170 MeV, which is more in accord with what is
found in lattice simulations of QCD with dynamical quarks.

\end{abstract}

\pacs{12.39.Fe, 12.38.Aw, 14.65.Bt}

\maketitle

\section{INTRODUCTION}

Euclidean-time hadronic current correlation functions contain
information about the hadronic spectrum, including information
about the temperature dependence of hadronic masses, the widths of
these states and the eventual disappearance of such status from
the spectrum at sufficiently high temperature. We have been
interested in recent calculations of spectral functions of
hadronic current correlators [\,1-3\,]. These calculations make
use of lattice results for the Euclidean-time correlation
functions, $G_P(\tau, T)$ and $G_V(\tau, T)$, of pseudoscalar and
vector currents. For example, for the pseudoscalar correlator, one
considers the equation \be G_P(\tau, T)=\int_0^\infty d \omega
\sigma_P(\omega, T) K(\tau, \omega, T)\,,\ee with \be K(\tau,
\omega, T)=\frac{\cosh[\omega(\tau-1/2T)]}{\sinh(\omega/2T)}\,,\ee
and attempts to obtain values of the spectral function
$\sigma_P(\omega, T)$ by using the maximum-entropy-method (MEM)
[\,4, 5\,]. (A similar procedure is used to obtain
$\sigma_V(\omega, T)$.) The MEM method is used, since the
inversion of Eq.\,(1.1) to obtain $\sigma_P(\omega, T)$ is an
"ill-conditioned problem"[\,2\,], given the limitations of the
lattice data.

The necessity of using such advanced numerical procedures is
discussed in Ref.\,[\,2\,]. As stated there, one has to introduce
assumptions as to the characteristics of the desired solutions
when using the MEM analysis. At this point, the physical
significance of the results for the spectral functions obtained
with the MEM procedure is uncertain in the case of resonant
structures with energies of several GeV found using that method.

Recently we have made some calculations of hadronic current
correlators and their associated spectral functions [\,6\,]. Once
we calculate the spectral functions using the Nambu-Jona-Lasinio
(NJL) model, we may obtain $G_P(\tau, T)$ and $G_V(\tau, T)$ by
using Eqs. (1.1) and (1.2). We may then compare our results for
these Euclidean-time correlators with those calculated using the
data obtained in the lattice simulations of QCD.

Since we have described our method of calculation of the spectral
functions in an earlier work [\,6\,], we here relegate that
material to the Appendix for ease of reference. A novel feature of
our work is use of temperature-dependent coupling parameters for
the NJL model. We have made use of such parameters in earlier
studies of meson properties at finite temperature and have
described the mesonic confinement-deconfinement transition
[\,7\,]. While the use of temperature-dependent coupling constants
for the NJL model is novel, we may note that the QCD coupling
constant $\alpha_s$ has been made temperature-dependent in a study
of the quark -gluon plasma equation of state in Ref. [\,8\,] and
[\,9\,]. (See Fig.\,1 of  Ref.\,[\,9\,].) Values of the pressure,
energy density and baryon density were presented as a function of
temperature and quite good fits to QCD lattice results were
obtained in Refs. [\,8\,] and [\,9\,].

In Ref.\,[\,6\,] we have introduced the spectral functions, \be
\sigma_P(\omega, T)=\frac{1}{\pi}\mbox{Im}C_P(\omega, T)\,,\ee and
\be \sigma_V(\omega, T)=\frac{1}{\pi}\mbox{Im}C_V(\omega, T)\,.\ee
The relation between these functions and the ones that appear in
the literature, $\overline{\sigma}_P(\omega, T)$ and
$\overline{\sigma}_V(\omega, T)$ [\,1-3\,], is given in the
Appendix. [\,See Eqs. (A30) and (A31).\,] In the Appendix we
describe the procedures used in the calculation of
$\mbox{Im}C_P(\omega, T)$ and $\mbox{Im}C_V(\omega, T)$.

The organization of our work is as follow. In Section II we
present some of the results obtained for the functions
$\mbox{Im}C_P(P_0, T)/P_0^2$ and $\mbox{Im}C_V(P_0, T)/P_0^2$. (We
identify $P_0$ with $\omega$, which is used in the literature for
the corresponding quantity.) In Section III and IV we compare the
values of $G_P(\tau, T)$ and $G_V(\tau, T)$ obtained in our
calculations and in the lattice simulations [\,2\,]. In those
sections we discuss the evidence for the temperature-dependent
coupling parameters we have used in other works [\,6, 7\,].
Finally, Section IV contains some additional discussion and
conclusions.

\section{Euclidean-time Correlators of Pseudoscalar Hadronic Currents}

The formalism reviewed in the Appendix is used to obtain the
results presented in this section. In Fig.1 we show the values of
$\mbox{Im} C_P(P_0, T)/P_0^2$ that were first presented in
Ref.\,[\,6\,]. The large peak at $T/T_c=1.2$ has its origin in the
properties of a pion-like mode that is present after deconfinement
has taken place. We note that no resonance is seen for $T/T_c=3.0$
[\,dot-dashed line\,], but resonance behavior is present for
$T/T_c=1.5$ [\,dashed line\,] and $T/T_c=2.0$ [\,dotted line\,].

 \begin{figure}
 \includegraphics[bb=0 0 280 235, angle=0, scale=1]{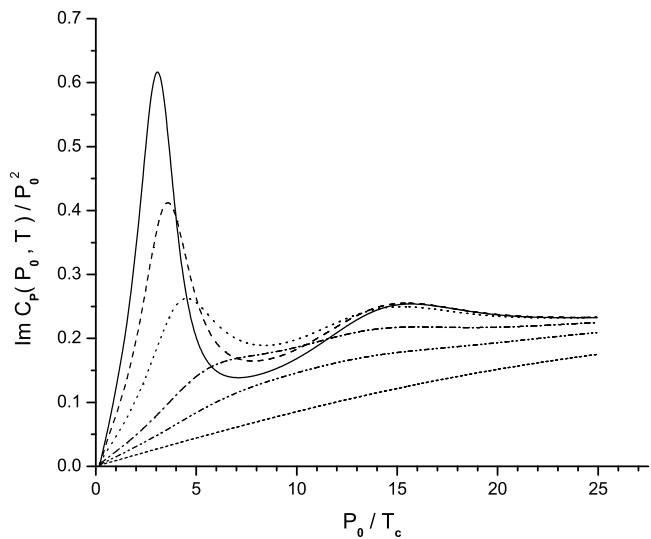}%
 \caption{Values of $\mbox{Im}\,C_P(P_0, T)/P_0^2$, obtained in Ref.\,[\,6\,], are shown for values of
 $T/T_c=1.2$ [\,solid line\,], $T/T_c=1.5$ [\,dashed line\,], $T/T_c=2.0$ [\,dotted line\,], $T/T_c=3.0$ [\,dot-dashed
 line\,], $T/T_c=4.0$ [\,double dot-dashed line\,], and $T/T_c=5.88$ [\,short dashed line\,]. Here
 we use $G_P(T)=G_P\,[\,1-0.17\,(T/T_c)\,]$ with $T_c=0.150$ GeV and $G_P=13.49\,\mbox{GeV}^{-2}$.}
 \end{figure}

Before we proceed, it is useful to discuss the properties of the
integral in Eq.\,(1.1). We wish to show that the calculation of
$G_P(\tau, T)$ is sensitive to the properties of $\sigma_P(\omega,
T)$ for relatively small $\omega$, when $\tau T \sim 0.5$. On the
other hand, when $\tau T$ is small (or near 1), the integral is
sensitive to the values of $\sigma_P(\tau, T)$ for large $\omega$,
where $\sigma_P(\tau, T)$ increases as $\omega^2$ and is largely
model independent. These features may be seen in Fig.2, where the
solid line represents our calculated values of $\sigma_P(\tau,
T)=(1/\pi)\mbox{Im} C_P(\omega, T)$ for $T=1.5\,T_c$. The dashed
line shows $K(\omega, \tau, T)$ for $\tau T = 0.5$, while the
dotted and dot-dashed lines show $K(\omega, \tau, T)$ for $\tau T
= 0.05$ and $\tau T = 0.10$, respectively. We note that the
$1/\omega$ singularity in $K(\omega, \tau, T)$ for small $\omega$
is compensated by the behavior of the spectral function at small
$\omega$, $\sigma_P(\omega, T) \sim \omega$. These remarks also
pertain to the calculation of $G_V(\tau, T)$ made using
$\sigma_V(\omega, T)$.

 \begin{figure}
 \includegraphics[bb=0 0 280 235, angle=0, scale=1]{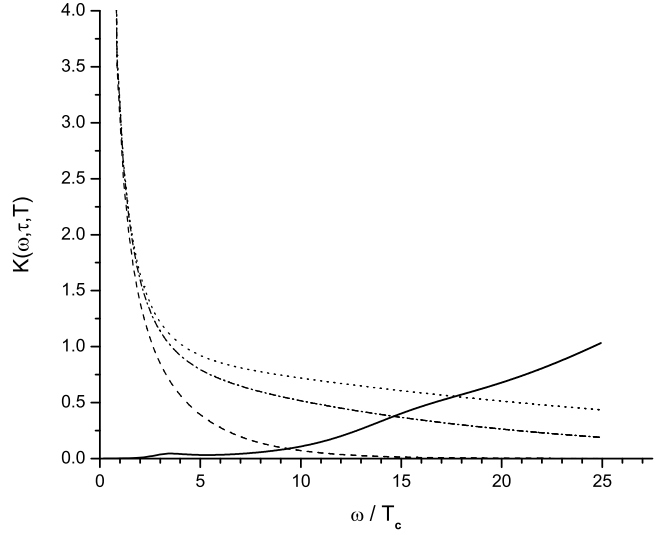}%
 \caption{Values of $K(\omega, \tau, T)$ are shown for $T=1.5\,T_c$ and $\tau T=0.05$ [\,dotted line\,],
 $\tau T=0.10$ [\,dot-dashed line\,], and $\tau T=0.50$ [\,dashed line\,]. The solid line represents
 $(1/\pi)\mbox{Im}C_P(\omega, T)$. Here, we use the notation $\omega=P_0$ and have put $T_c=0.150\,\mbox{GeV}$.}
 \end{figure}

 \begin{figure}
 \includegraphics[bb=0 0 280 235, angle=0, scale=1]{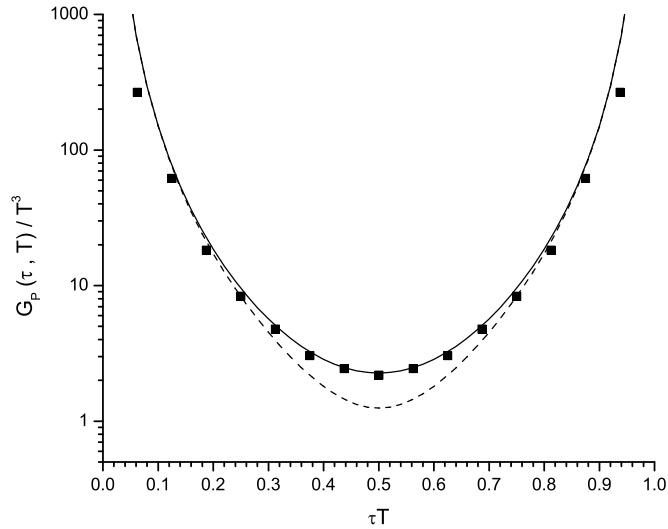}%
 \caption{Values of $G_P(\tau, T)/T^3$ are shown as a function of $\tau T$, with $T=1.5\,T_c$. The
 solid line represents the result of our calculation made for $G_P(T)=G_P\,[\,1-0.17\,(T/T_c)\,]$
 with $G_P=13.49\,\mbox{GeV}^{-2}$. The dotted line is obtained when we use a constant value $G_P(T)=G_P$.
 The data (squares) are taken from Ref.\,[\,2\,] for the case $T=1.5\,T_c$. }
 \end{figure}

 \begin{figure}
 \includegraphics[bb=0 0 280 235, angle=0, scale=1]{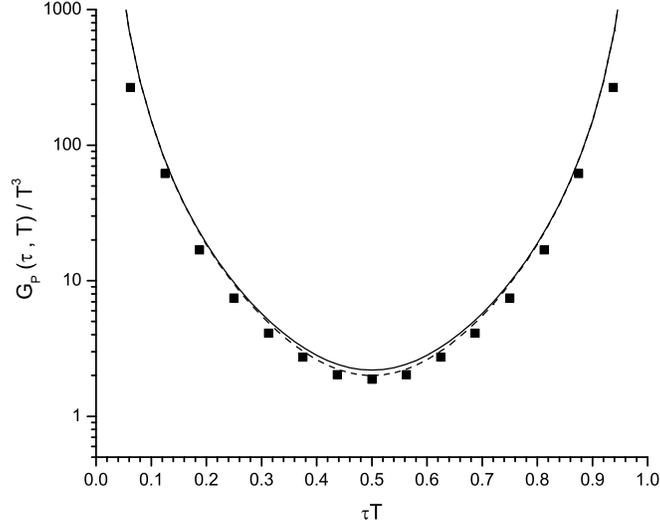}%
 \caption{Values of $G_P(\tau, T)/T^3$ are shown as a function of $\tau T$, for $T=3.0\,T_c$. The
 solid line is the result of our calculation made for $G_P(T)=G_P\,[\,1-0.17\,(T/T_c)\,]$
 with $G_P=13.49\,\mbox{GeV}^{-2}$, while the dotted line is obtained when we put $G_P(T)=G_P$.
 The data (squares) are taken from Ref.\,[\,2\,] for the case $T=1.5\,T_c$.}
 \end{figure}

 \begin{figure}
 \includegraphics[bb=0 0 280 235, angle=0, scale=1]{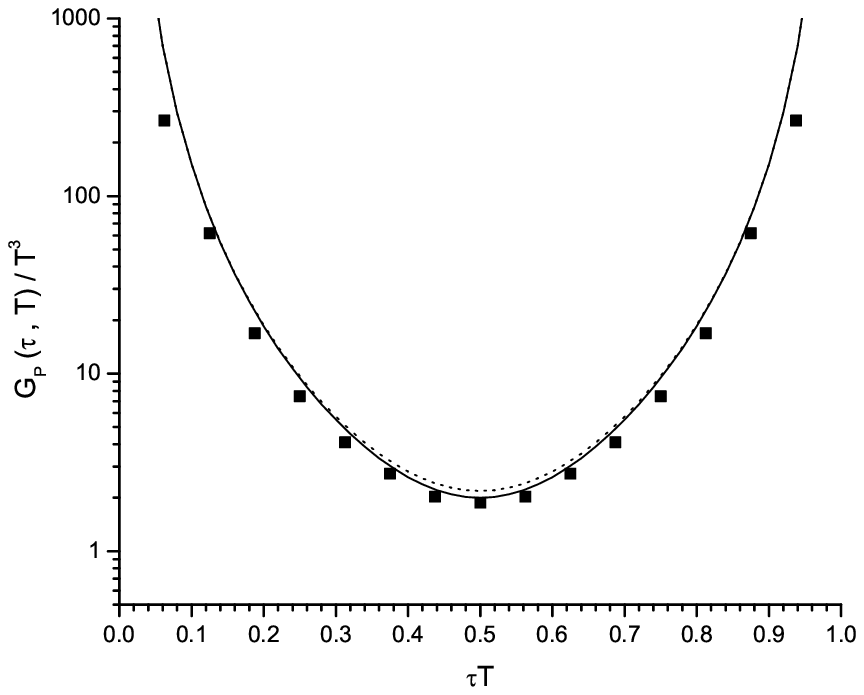}%
 \caption{Values of $G_P(\tau, T)/T^3$  are shown for $T=3.0\,T_c$. Here the
 solid line is the same as the dashed line in Fig.\,4 and corresponds to $G_P=0$. The dotted line is obtained
 when we use the constant value $G_P(T)=G_P=13.49\,\mbox{GeV}^{-2}$. }
 \end{figure}

In Fig.\,3 we show the data we have taken from Ref.\,[\,2\,] as
squares. The solid line shows the result of our calculation of
$G_P(\tau, T)/T^3$, as obtained from our calculated values of
$\sigma_P(\omega, T)$ at $T=1.5\,T_c$. The deviation of our
calculation from the data for $\tau T \simeq 0$ or $\tau T \simeq
1$ is inconsequential, given our remarks made in the previous
paragraph. We also show the result for the case $G_P(T)=G_P=13.49
\,\mbox{GeV}^{-2}$ as a dotted curve. When comparing the solid
curve and the dotted curve we see evidence for the temperature
dependence we have used: $G_P(T)=G_P[\,1-0.17(T/T_c)\,]$. (We have
obtained the value $G_P=13.49 \,\mbox{GeV}^{-2}$ by calculating
the pion mass at $T=0$ in our model.)

In Fig.\,4 we show, as a solid line, our result for $G_P(\tau,
T)/T^3$ at $T=3.0\,T_c$. In this case the fit to the data taken
from Ref.\,[\,2\,] is poor. Therefore, we have investigated the
case in which $G_P(T)=0$ and show the result as a dashed line in
Fig.\,4. The fit to the data of Ref.\,[\,2\,] is improved
somewhat. We may suggest that the suppression of $G_P(T)$ at large
temperatures may be greater than that given by the form we have
used, $G_P(T)=G_P[\,1-0.17(T/T_c)\,]$.

In Fig.\,5 we again consider the value $T=3.0\,T_c$. In this case,
the solid line is the result obtained when $G_P(T)=0$, while the
dotted line results when we use a constant value for
$G_P(T)=G_P=13.49 \,\mbox{GeV}^{-2}$. We may compare our results
for the spectral functions to those obtained using the MEM
procedure and depicted in Ref.\,[\,2\,]. One essential difference
is that we do not see the resonant behavior reported there for
$T=3.0\,T_c$ at a rather high energy of about 2.4 GeV. The peak in
$\overline{\sigma}_P(\omega, T)/\omega^2$ for $T=1.5\,T_c$ is at
about 1 GeV in Ref.\,[\,2\,], while for $T=1.5\,T_c$ our peak in
Fig. 1 is at about 0.55 GeV.

In estimating the energies of the peaks we have used $T_c=150$
MeV. However, if we use $T_c=270$ MeV, which is more appropriate
for a lattice calculation without dynamical quarks, the resonant
structures would appear at still higher energy.

\section{Euclidean-time Correlators of Lorentz-vector Hadronic Currents}

 \begin{figure}
 \includegraphics[bb=0 0 280 235, angle=0, scale=1]{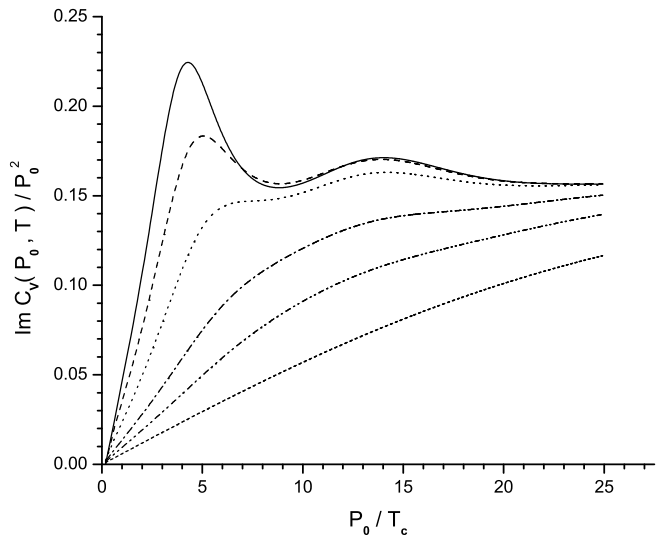}%
 \caption{Values of $\mbox{Im}\,C_V(P_0, T)/P_0^2$, obtained in Ref.\,[\,6\,], are shown for values of
 $T/T_c=1.2$ [\,solid line\,], $T/T_c=1.5$ [\,dashed line\,], $T/T_c=2.0$ [\,dotted line\,], $T/T_c=3.0$ [\,dot-dashed
 line\,], $T/T_c=4.0$ [\,double dot-dashed line\,], and $T/T_c=5.88$ [\,short dashed line\,]. Here
 we use $G_V(T)=G_V\,[1-0.17\,(T/T_c)]$ with $T_c=0.150$ GeV and $G_V=11.46\,\mbox{GeV}^{-2}$}
 \end{figure}

 \begin{figure}
 \includegraphics[bb=0 0 280 235, angle=0, scale=1]{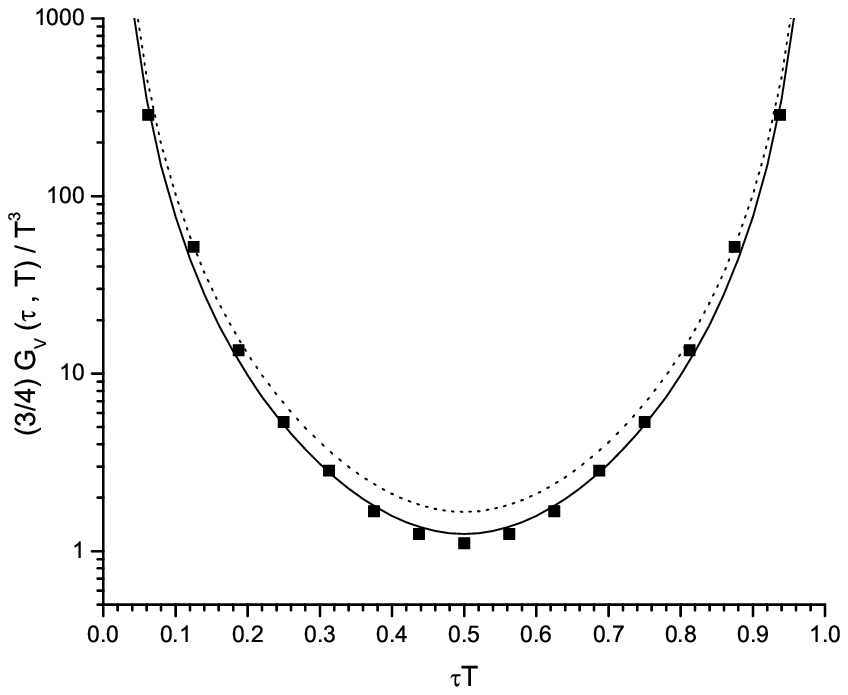}%
 \caption{Values of $(3/4)G_V(\tau, T)/T^3$ are shown for $T=1.5\,T_c$. Here, the
 solid line represents the result when we use $G_V(T)=G_V\,[\,1-0.17\,(T/T_c)\,]$
 with $G_V=11.46\,\mbox{GeV}^{-2}$. The dotted line is obtained when we use a constant value of
 $G_V(T)=G_V=11.46\,\mbox{GeV}^{-2}$. The data (squares) are taken from Ref.\,[\,2\,] for the case $T=1.5\,T_c$.}
 \end{figure}

 \begin{figure}
 \includegraphics[bb=0 0 280 235, angle=0, scale=1]{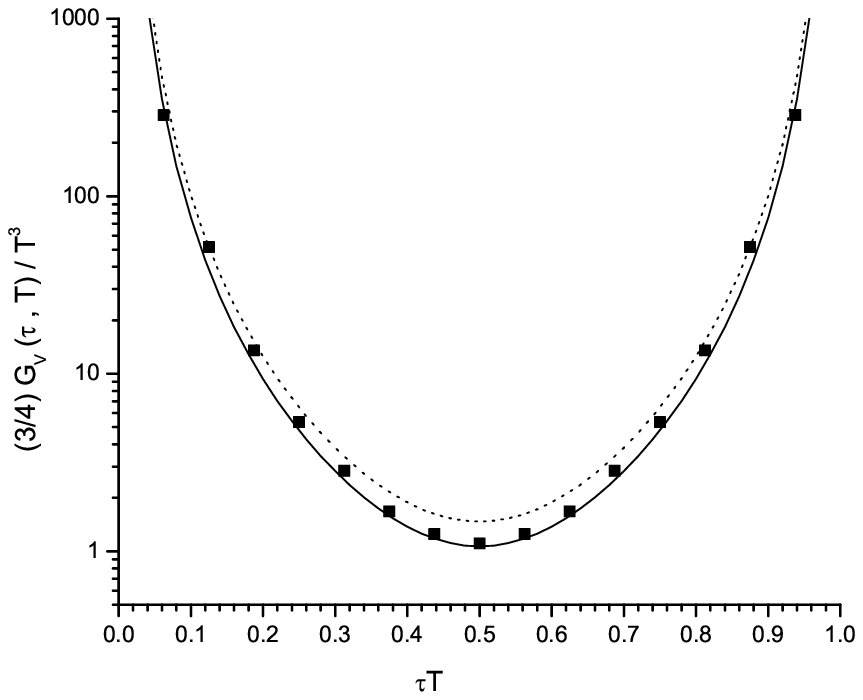}%
 \caption{Values of $(3/4)G_V(\tau, T)/T^3$ are shown for $T=3.0\,T_c$. Here, the
 solid line represents the result when we use $G_V(T)=G_V\,[\,1-0.17\,(T/T_c)\,]$
 with $G_V=11.46\,\mbox{GeV}^{-2}$. The dotted line is obtained when we use a constant value of
 $G_V(T)=G_V=11.46\,\mbox{GeV}^{-2}$. The data (squares) are taken from Ref.\,[\,2\,] for the case $T=3.0\,T_c$.}
 \end{figure}

We again refer to the Appendix for a review of our procedures for
the calculation of the spectral functions from which we may obtain
the Euclidean-time correlator using Eq.\,(1.1), with
$\sigma_V(\omega, T)$ replacing $\sigma_P(\omega, T)$. In Fig. 6
we present our values of $\mbox{Im} C_V(P_0, T)/P_0^2$, which may
be compared to the values of $\overline{\sigma}_V(\omega,
T)/\omega^2$ given in Fig. 1 of Ref.\,[\,2\,]. In Ref.\,[\,2\,] we
find a peak at about 1 GeV for $T=1.5\,T_c$ and at about 2.5 GeV
for $T=3.0\,T_c$. On the other hand, there is some weak resonant
behavior seen for $T=1.5\,T_c$ [\,dashed line\,] in our Fig. 6 at
about 0.75 GeV which may reflect a residual enhancement due to a
$\rho$-like mode that is present after the
confinement-deconfinement transition has taken place. At
$T=3.0\,T_c$ [\,dot-dashed line\,], we see no resonance
enhancement in our work, in contrast to what is obtained by the
MEM analysis of Ref.\,[\,2\,].

In Fig.\,7 we show the data taken from Ref.\,[\,2\,] as squares.
Here, there is hardly any difference seen in the data reported for
$T=1.5\,T_c$ and $T=3.0\,T_c$. In Fig. 7, the solid line
represents our results for $(3/4)G_V(\tau, T)/T^3$ at
$T=1.5\,T_c$. We only achieve a fair fit to the data, but the fit
is decidedly better than that obtained when a constant value of
$G_V(T)=G_V=11.46\,\mbox{GeV}^{-2}$ is used [\,dotted line\,] (See
the appendix for a discussion of the factor 3/4 used when making
comparison to the data of Ref.\,[\,2\,] in the case of vector
current correlators.)

In Fig. 8 we compare our result for $(3/4)G_V(\tau, T)/T^3$ at
$T=3.0\,T_c$ [\,solid line\,] with the data of Ref.\,[\,2\,]. Here
the fit is better than that of Fig.\,4. Again, the use of a
constant value of $G_V(T)=G_V=11.46\,\mbox{GeV}^{-2}$ yields a
poor result. We suggest that our analysis tends to support our
choice of temperature-dependent coupling parameters for the NJL
model.

\section{discussion}

It is of interest to obtain further insight into the results shown
in Figs.\,4 and 5. To that end, we show various calculations made
for $\mbox{Im}C_P(P_0, T)/P_0^2$ in Fig.\,9. There, for
$T=3.0\,T_c$, the solid line is the result of our model, the
dotted curve corresponds to the use of a constant value of the
coupling parameter $G_P(T)=G_P=13.49\,\mbox{GeV}^{-2}$, while the
dashed line is the result for $G_P(T)=0$. A comparison of the
solid curve and the dashed curve leads to some understanding of
the results shown in Fig.\,4, while a comparison of the dotted
curve and the dashed curve leads to further understanding of the
results shown in Fig.\,5. Similar results are given for
$\mbox{Im}C_V(P_0, T)/P_0^2$ in Fig. 10. For that figure, a
comparison of the solid curve and the dashed curve gives some
insight into the results shown in Fig. 8, where the dotted line
corresponds to $G_V(T)=G_V=11.46\,\mbox{GeV}^{-2}$ and the solid
curve represents the results of our model.

We now return to the pseudoscalar case for $T=1.5\,T_c$. In
Fig.\,11 we show $\mbox{Im}C_P(P_0, T)/P_0^2$ for our model
[\,solid line\,], for $G_P(T)=0$ [dashed line], and for the case
$G_P(T)=G_P=13.49\,\mbox{GeV}^{-2}$. We recall that our model,
with the temperature-dependent coupling parameter, gave rise to a
excellent fit to the data, as seen in Fig. 3. It is also of
interest to present values of $\sigma_P(\omega,
T)=(1/\pi)\mbox{Im}C_P(\omega, T)/\omega^2$ for $T=1.5\,T_c$. In
Fig.\,12, the result of our model is shown as a solid line, the
dot-dashed line is for $G_P(T)=0$, and the dotted line is obtained
when $G_P(T)=G_P=13.49\,\mbox{GeV}^{-2}$. It is seen, that for
small values of $P_0$, on the whole, the dotted line lies below
the other curves, giving rise to the behavior seen in Fig.\,3 for
$\tau T \simeq 0.5$.

It is worth mentioning that we have some additional evidence of
the utility of temperature-dependent coupling parameters for the
NJL model. We note that the confinement-deconfinement transition
takes place in the range $150\,\mbox{MeV} \leq T_c \leq
170\,\mbox{MeV}$ for QCD with dynamical quarks. We then inspect
Fig.\,5 of Ref.\,[\,10\,], where the constituent quark mass of the
NJL model is presented as a function of temperature for the case
of a temperature-independent coupling constant. The mass value is
$330\,\mbox{MeV}$ at $T=0$ and is about $260-300\,\mbox{MeV}$ when
$150\,\mbox{MeV}\leq T \leq 170\,\mbox{MeV}$. Thus, we do not see
the (partial) restoration of the chiral symmetry that is expected
for $T \sim T_c$. On the other hand, we see in Fig.\,1 of
Ref.\,[\,7\,], where we have used a temperature-dependent coupling
parameter, that we have $m_u=364\,\mbox{MeV}$ at $T=0$ and
$m_u(T)$ in the range of 50 to $100\,\mbox{MeV}$ for
$150\,\mbox{MeV}$ $\leq T \leq 170\,\mbox{MeV}$. That is much more
in accord with the (partial) restoration of chiral symmetry when
$T \sim T_c$. If we wish to consider the NJL as a useful
low-energy model of QCD, it is much easier to discuss the
confinement-deconfinement transition if we use
temperature-dependent coupling parameters.

 \begin{figure}
 \includegraphics[bb=0 0 280 235, angle=0, scale=1]{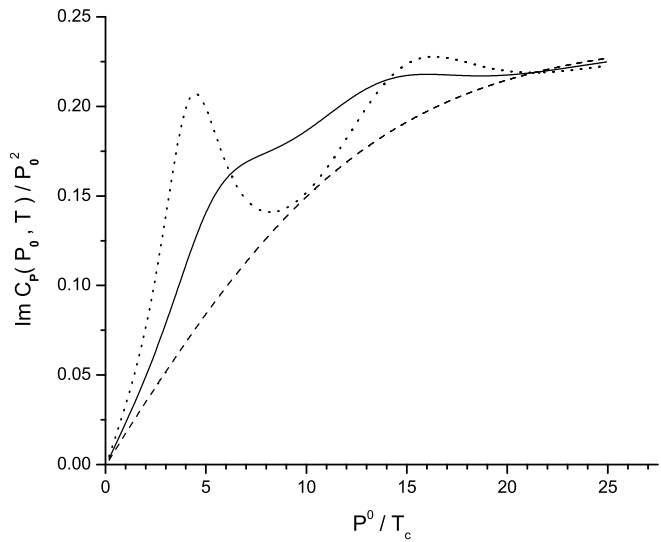}%
 \caption{Values of $\mbox{Im}\,C_P(P_0, T)/P_0^2$ are shown for $T=3.0\,T_c$. The solid line is the result of our model
 with temperature-dependent coupling parameters, the dotted line is obtained in the absence of the temperature dependence
 ($G_P(T)=G_P=13.49\,\mbox{GeV}^{-2}$), and the dashed line represents the result for $G_P(T)=0$. }
 \end{figure}

 \begin{figure}
 \includegraphics[bb=0 0 280 235, angle=0, scale=1]{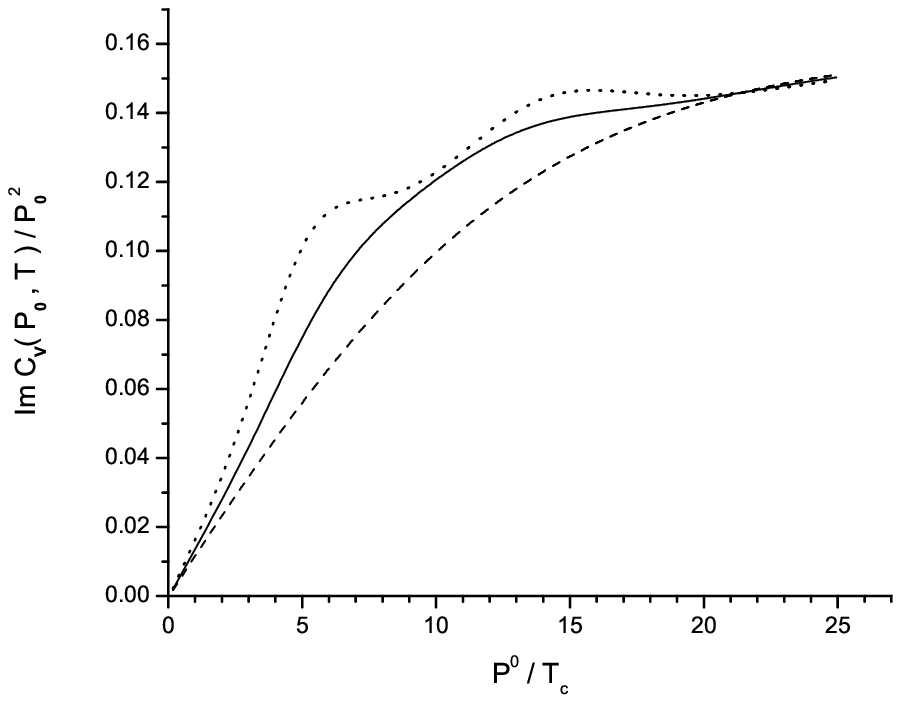}%
 \caption{Values of $\mbox{Im}\,C_V(P_0, T)/P_0^2$ are shown for $T=3.0\,T_c$. The solid line is the result of our model
 with temperature-dependent coupling parameters, the dotted line is obtained in the absence of the temperature dependence
 ($G_V(T)=G_V=11.46\,\mbox{GeV}^{-2}$), and the dashed line represents the result for $G_V(T)=0$. }
 \end{figure}

 \begin{figure}
 \includegraphics[bb=0 0 280 235, angle=0, scale=1]{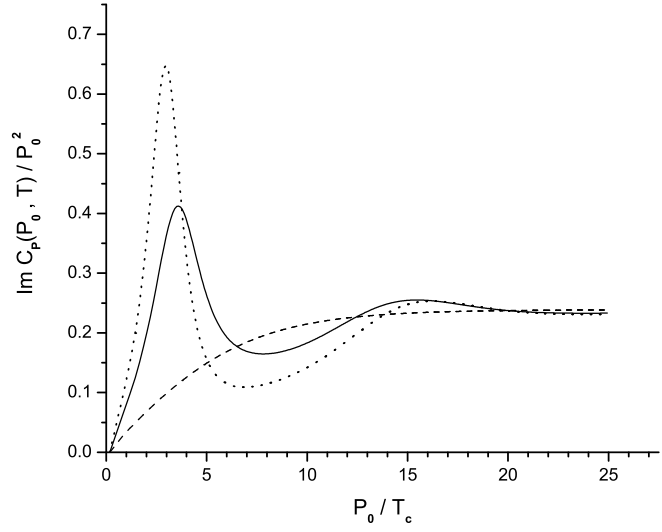}%
 \caption{Values of $\mbox{Im}\,C_P(P_0, T)/P_0^2$ are shown for $T=1.5\,T_c$. Here,
 the solid line is the result of our model, the dashed line represent the result for $G_P(T)=0$,
 while the dotted line is obtained when we use $G_P(T)=G_P=13.49\,\mbox{GeV}^{-2}$. [\,See Fig.\,3 for the values of $G_P(\tau,
 T)/T^3$ calculated for $T=1.5\,T_c$, using the values of $\mbox{Im}\,C_P(P_0, T)/P_0^2$ shown here as the solid and dotted
 line.\,]}
 \end{figure}

 \begin{figure}
 \includegraphics[bb=0 0 280 235, angle=0, scale=1]{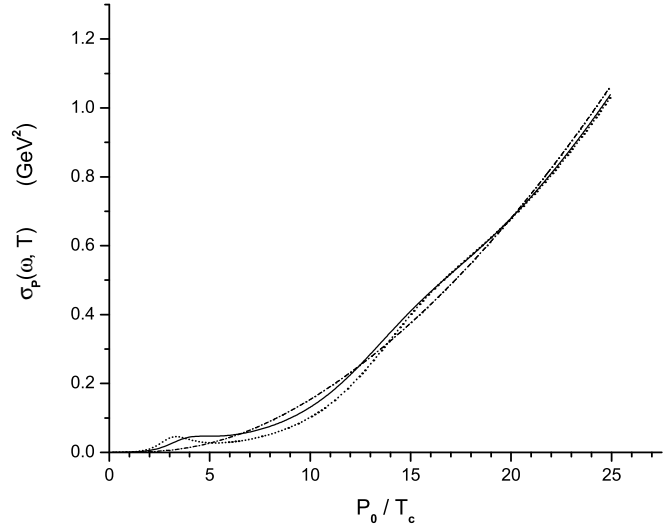}%
 \caption{Values of $\sigma_P(\omega, T)=(1/\pi)\mbox{Im}\,C_P(P_0, T)$ are shown for $T=1.5\,T_c$.
 Here, the solid line corresponds to our model, with $G_P(T)=G_P\,[\,1-0.17\,(T/T_c)\,]$, the dot-dashed line
 is obtained when $G_P(T)=0$, and the dashed line is for the case $G_P(T)=G_P=13.49\,\mbox{GeV}^{-2}$. (See Fig.\,13.)}
 \end{figure}

 \begin{figure}
 \includegraphics[bb=0 0 280 235, angle=0, scale=1]{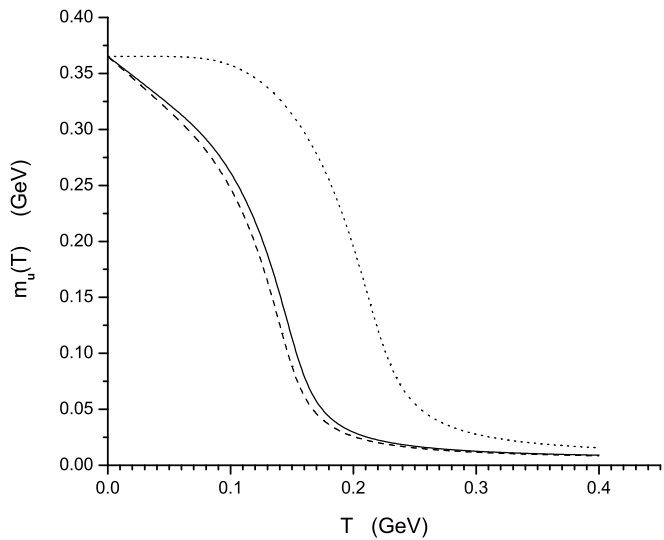}%
 \caption{We exhibit the values of $m_u(T)$ obtained using Eq.\,(5.38) of Ref.\,[11], with
 $m^0=5.50\,\mbox{MeV}$ and $\Lambda=0.631\,\mbox{GeV}$. The dotted curve corresponds to the use of a
 constant value $G=5.691\,\mbox{GeV}^{-2}$, in the notation of Ref.\,[11]. For the solid and dashed
 curves we have used $G(T)=G\,[\,1-0.17(T/T_c)\,]$. For the solid curve we have put
 $T_c=0.170\,\mbox{GeV}$, while for the dashed curve, we have used $T_c=0.150\,\mbox{GeV}$ in our
 parametrization of $G(T)$.
 }
 \end{figure}

 \begin{figure}
 \includegraphics[bb=0 0 280 235, angle=0, scale=1]{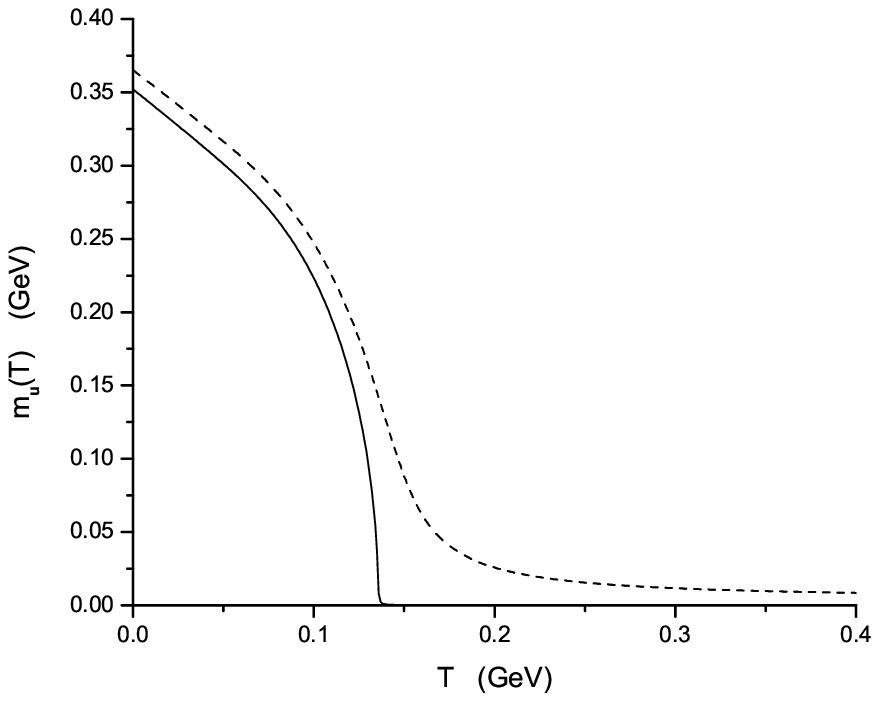}%
 \caption{Values of $m_u(T)$ are shown. The dashed curve is calculated with $m^0=5.50\,\mbox{MeV}$.
 Here, $G(T)=G\,[\,1-0.17\,(T/T_c)\,]$, with $G=5.691\,\mbox{GeV}^{-2}$ and $T_c=0.150\,\mbox{GeV}$. The
 solid curve is calculated with the same value of $G(T)$ and $T_c$, but with $m^0=0$. From the solid
 curve, we see that chiral symmetry is restored at $T=0.136\,\mbox{GeV}$ when $m^0=0$.
 }
 \end{figure}

For ease of reference,we have calculated the constituent mass of
the up quark using the equation for the temperature-dependent
constituent mass given in Ref.\,[\,11\,]. We have used a current
quark mass of $m^0=5.5\,\mbox{MeV}$ and a momentum cutoff of
$\Lambda=0.631\,\mbox{GeV}$. In Fig.\,13, the dotted curve shows
the result obtained with $G=5.691\,\mbox{GeV}^{-2}$ (in the
notation of Ref.\,[\,11\,]). From Fig.\,13 we see that at
$T=0.150\,\mbox{GeV}$, $m_u=0.318\,\mbox{GeV}$, while at
$T=0.170\,\mbox{GeV}$, $m_u=0.276\,\mbox{GeV}$. The dashed and
solid curves represent the result when
$G(T)=G[\,1-0.17(T/T_c)\,]$. For the solid curve
($T_c=0.170\,\mbox{GeV}$), $m_u=77\,\mbox{MeV}$ at
$T=0.170\,\mbox{GeV}$. For the dashed curve
($T_c=0.150\,\mbox{GeV}$), $m_u=54\,\mbox{MeV}$ at
$T=0.150\,\mbox{GeV}$. Again, we see that it is much easier to
discuss the (partial) restoration of chiral symmetry at the
confinement-deconfinement transition when we use the
temperature-dependent coupling parameters of our model.

It is also of interest to exhibit the role played by the current
quark mass. In Fig.\,14, the dashed curve, which was calculated
for $T_c=0.150\,\mbox{GeV}$, is the same as the dashed curve in
Fig.\,13. In Fig.\,14, the solid curve shows the result when
$m^0=0$. Here, we see restoration of chiral symmetry at
$T=0.136\,\mbox{GeV}$, when $G(T)=G[\,1-0.17(T/T_c)\,]$ with
$T_c=0.150\,\mbox{GeV}$.

We note that the value of $T_c=0.136\,\mbox{GeV}$ is still within
the uncertainty of the transition temperature for three-flavor
QCD. For example, in Ref.\,[\,12\,] the transition temperatures
are given for two-flavor and three-flavor QCD. In the latter case,
$T_c=154 \pm 8\,\mbox{MeV}$, with a suggested systematic error
similar to the statistical error, so that $T_c=154 \pm 8 \pm
8\,\mbox{MeV}$ [\,12\,]. (We have used coupling constants
determined in our studies of the three-flavor NJL model, so that
consideration of the transition temperature for that case is
appropriate.)

If we wish to assign a physical interpretation of the parameter
$T_c$ in the expression for $G(T)$, we may use
$G(T)=G[\,1-0.135(T/T_c)\,]$ with $T_c=0.150\,\mbox{GeV}$. That
choice gives rise to restoration of chiral symmetry at
$T=T_c=0.150\,\mbox{GeV}$ in the NJL model with $m^0=0$. If we
maintain the value $m^0=0$, but use a constant value for
$G(T)=5.691 \,\mbox{GeV}^{-2}$, with $\Lambda=0.631\,\mbox{GeV}$,
we find restoration of chiral symmetry at $T_c=208\,\mbox{MeV}$.

The calculations reported in Fig. 13 and 14 may also be made using
our Gaussian cutoff, $\mbox{exp}(-\overrightarrow{k}^2/\alpha^2)$,
with $\alpha=0.605\,\mbox{GeV}$. For example, we may consider the
dashed curve of Fig.\,14. If we use the Gaussian cutoff and use
$G=6.004\,\mbox{GeV}^{-2}$, instead of $G=5.691\,\mbox{GeV}^{-2}$,
we obtain a curve that is very close to the dashed curve of
Fig.\,14 for $T>0.080\,\mbox{GeV}$. However, the mass at $T=0$ is
$382\,\mbox{MeV}$ instead of $364\,\mbox{MeV}$ which was the value
obtained using the sharp cutoff of $\Lambda=0.631\,\mbox{GeV}$ and
$G=5.691\,\mbox{GeV}^{-2}$.

We have performed what are essentially parameter-free calculations
of hadronic spectral functions and have computed the corresponding
Euclidean-time correlation functions. The values of the coupling
parameters, $G_S$ and $G_V$, were fixed in calculations of meson
properties at $T=0$. We have used the temperature dependence,
$G(T)=G[\,1-0.17(T/T_c)\,]$, which was introduced in earlier work
in which we studied the (mesonic) confinement-deconfinement
transition [\,7\,]. We believe it is of interest to see that we
obtain reasonable values for the Euclidean correlators. Of more
importance, however, is our observation that we find some evidence
for the temperature dependence of the NJL coupling parameters that
we have used in other works. By analogy, we expect that the
coupling parameters should also be density-dependent, and we have
introduced such density dependence in earlier work [\,13\,].
Density dependence of the coupling parameters may be particularly
important, given the strong interest in diquark condensates and
color superconductivity at high baryon density [\,14\,].

\appendix
  \renewcommand{\theequation}{A\arabic{equation}}
  \setcounter{equation}{0}  
  \section*{APPENDIX}  

For ease of reference, we present a discussion of our calculation
of hadronic current correlators taken from Ref.\,[\,6\,]. The
procedure we adopt is based upon the real-time finite-temperature
formalism, in which the imaginary part of the polarization
function may be calculated. Then, the real part of the function is
obtained using a dispersion relation. The result we need for this
work has been already given in the work of Kobes and Semenoff
[\,15\,]. (In Ref.\,[\,15\,] the quark momentum in Fig.\,2 is $k$
and the antiquark momentum is $k-P$. We will adopt that notation
in this section for ease of reference to the results presented in
Ref.\,[\,15\,].) With reference to Eq.\,(5.4) of Ref.\,[\,15\,],
we write the imaginary part of the scalar polarization function as
\be \mbox{Im}\,J_S(\textit{P}\,{}^2,
T)=\frac12(2N_c)\beta_S\,\epsilon(\textit{P}\,{}^0)\mytint
ke^{-\vec
k\,{}^2/\alpha^2}\left(\frac{2\pi}{2E_1(k)2E_2(k)}\right)\\\nonumber
\{(1-n_1(k)-n_2(k))
\delta(\textit{P}\,{}^0-E_1(k)-E_2(k))\\\nonumber-(n_1(k)-n_2(k))
\delta(\textit{P}\,{}^0+E_1(k)-E_2(k))\\\nonumber-(n_2(k)-n_1(k))
\delta(\textit{P}\,{}^0-E_1(k)+E_2(k))\\\nonumber-(1-n_1(k)-n_2(k))
\delta(\textit{P}\,{}^0+E_1(k)+E_2(k))\}\,.\ee Here,
$E_1(k)=[\,\vec k\,{}^2+m_1^2(T)\,]^{1/2}$. Relative to Eq.\,(5.4)
of Ref.\,[\,15\,], we have changed the sign, removed a factor of
$g^2$ and have included a statistical factor of $2N_c$, where the
factor of 2 arises from the flavor trace. In addition, we have
included a Gaussian regulator, $\exp[\,-\vec k\,{}^2/\alpha^2\,]$,
with $\alpha=0.605$ GeV, which is the same as that used in most of
our applications of the NJL model in the calculation of meson
properties. We also note that \be n_1(k)=\frac1{e^{\,\beta
E_1(k)}+1}\,,\ee and \be n_2(k)=\frac1{e^{\,\beta E_2(k)}+1}\,.\ee
For the calculation of the imaginary part of the polarization
function, we may put $\ksq=m_1^2(T)$ and $(k-P)^2=m_2^2(T)$, since
in that calculation the quark and antiquark are on-mass-shell. In
Eq.\,(A1) the factor $\beta_S$ arises from a trace involving Dirac
matrices, such that
\be \beta_S&=&-\mbox{Tr}[\,(\slr k+m_1)(\slr k-\slr P+m_2)\,]\\
&=&2P^2-2(m_1+m_2)^2\,,\ee where $m_1$ and $m_2$ depend upon
temperature. In the frame where $\vec P=0$, and in the case
$m_1=m_2$, we have $\beta_S=2P_0^2(1-{4m^2}/{P_0^2})$. For the
scalar case, with $m_1=m_2$, we find \be \mbox{Im}\,J_S(P^2,
T)=\frac{N_cP_0^2}{4\pi}\left(1-\frac{4m^2(T)}{P_0^2}\right)^{3/2}
e^{-\vec k\,{}^2/\alpha^2}[\,1-2n_1(k)\,]\,,\ee where \be \vec
k\,{}^2=\frac{P_0^2}4-m^2(T)\,.\ee


For pseudoscalar mesons, we replace $\beta_S$ by
\be \beta_P&=&-\mbox{Tr}[\,i\gamma_5(\slr k+m_1)i\gamma_5(\slr k-\slr P+m_2)\,]\\
&=&2P^2-2(m_1-m_2)^2\,,\ee which for $m_1=m_2$ is $\beta_P=2P_0^2$
in the frame where $\vec P=0$. We find, for the $\pi$ mesons, \be
\mbox{Im}\,J_P(P^2,T)=\frac{N_cP_0^2}{4\pi}\left(1-\frac{4m^2(T)}{P_0^2}\right)^{1/2}
e^{-\vec k\,{}^2/\alpha^2}[\,1-2n_1(k)\,]\,,\ee where $ \vec
k\,{}^2={P_0^2}/4-m_u^2(T)$, as above. Thus, we see that, relative
to the scalar case, the phase space factor has an exponent of 1/2
corresponding to a \textit{s}-wave amplitude. For the scalars, the
exponent of the phase-space factor is 3/2, as seen in Eq.\,(A6).

For a study of vector mesons we consider \be
\beta_{\mu\nu}^V=\mbox{Tr}[\,\gamma_\mu(\slr k+m_1)\gamma_\nu(\slr
k-\slr P+m_2)\,]\,,\ee and calculate \be
g^{\mu\nu}\beta_{\mu\nu}^V=4[\,P^2-m_1^2-m_2^2+4m_1m_2\,]\,,\ee
which, in the equal-mass case, is equal to $4P_0^2+8m^2(T)$, when
$m_1=m_2$ and $\vec P=0$. This result will be needed when we
calculate the correlator of vector currents in the next section.
Note that, for the elevated temperatures considered in this work,
$m_u(T)=m_d(T)$ is quite small, so that $4P_0^2+8m_u^2(T)$ can be
approximated by $4P_0^2$, when we consider the vector current
correlation functions. In that case, we have \be
\mbox{Im}\,J_V(P^2,T) \simeq
\frac{2}{3}\mbox{Im}\,J_P(P^2,T)\,.\ee At this point it is useful
to define functions that do not contain that Gaussian regulator:
\be\mbox{Im}\,\tilde{J}_P(P^2,T)=\frac{N_cP_0^2}{4\pi}\left(1-\frac{4m^2(T)}{P_0^2}\right)^{1/2}[\,1-2n_1(k)\,]\,,\ee
and
\be\mbox{Im}\,\tilde{J}_V(P^2,T)=\frac{2}{3}\frac{N_cP_0^2}{4\pi}\left(1-\frac{4m^2(T)}{P_0^2}\right)^{1/2}[\,1-2n_1(k)\,]\,,\ee
For the functions defined in Eq.\,(A14) and (A15) we need to use a
twice-subtracted dispersion relation to obtain
$\mbox{Re}\,\tilde{J}_P(P^2,T)$, or
$\mbox{Re}\,\tilde{J}_V(P^2,T)$. For example,
\be\mbox{Re}\,\tilde{J}_P(P^2,T)=\mbox{Re}\,\tilde{J}_P(0,T)+
\frac{P^2}{P_0^2}[\,\mbox{Re}\,\tilde{J}_P(P_0^2,T)-\mbox{Re}\,\tilde{J}_P(0,T)\,]+\\\nonumber
\frac{P^2(P^2-P_0^2)}{\pi}\int_{4m^2(T)}^{\tilde{\Lambda}^{2}}
ds\frac{\mbox{Im}\,\tilde{J}_P(s,T)}{s(P^2-s)(P_0^2-s)}\,,\ee
where $\tilde{\Lambda}^{2}$ can be quite large, since the integral
over the imaginary part of the polarization function is now
convergent. We may introduce $\tilde{J}_P(P^2,T)$ and
$\tilde{J}_V(P^2,T)$ as complex functions, since we now have both
the real and imaginary parts of these functions. We note that the
construction of either $\mbox{Re}\,J_P(P^2,T)$, or
$\mbox{Re}\,J_V(P^2,T)$, by means of a dispersion relation does
not require a subtraction. We use these functions to define the
complex functions $J_P(P^2,T)$ and $J_V(P^2,T)$.

In order to make use of Eq.\,(A16), we need to specify
$\tilde{J}_P(0)$ and $\tilde{J}_P(P_0^2)$. We found it useful to
take $P_0^2=-1.0$ \gev2 and to put $\tilde{J}_P(0)=J_P(0)$ and
$\tilde{J}_P(P_0^2)=J_P(P_0^2)$. The quantities $\tilde{J}_V(0)$
and $\tilde{J}_V(P_0^2)$ are determined in an analogous function.
This procedure in which we fix the behavior of a function such as
$\mbox{Re}\tilde{J}_V(P^2)$ or $\mbox{Re}\tilde{J}_V(P^2)$ is
quite analogous to the procedure used in Ref.\,[\,16\,]. In that
work we made use of dispersion relations to construct a continuous
vector-isovector current correlation function which had the
correct perturbative behavior for large $P^2\rightarrow-\infty$
and also described that low-energy resonance present in the
correlator due to the excitation of the $\rho$ meson. In
Ref.\,[\,16\,] the NJL model was shown to provide a quite
satisfactory description of the low-energy resonant behavior of
the vector-isovector correlation function.

We now consider the calculation of temperature-dependent hadronic
current correlation functions. The general form of the correlator
is a transform of a time-ordered product of currents, \be iC(P^2,
T)=\int d^4xe^{iP\cdot x}<\!\!<T(j(x)j(0))>\!\!>\,,\ee where the
double bracket is a reminder that we are considering the finite
temperature case.

For the study of pseudoscalar states, we may consider currents of
the form $j_{P,i}(x)=\tilde{q}(x)i\gamma_5\lambda^iq(x)$, where,
in the case of the $\pi$ mesons, $i=1,2$ and $3$. For the study of
scalar-isoscalar mesons, we introduce
$j_{S,i}(x)=\tilde{q}(x)\lambda^i q(x)$, where $i=0$ for the
flavor-singlet current and $i=8$ for the flavor-octet current
[\,7\,].

In the case of the pseudoscalar-isovector mesons, the correlator
may be expressed in terms of the basic vacuum polarization
function of the NJL model, $J_P(P^2, T)$ [\,11, 17, 18\,]. Thus,
\be C_P(P^2, T)=J_P(P^2, T)\frac{1}{1-G_{P}(T)J_P(P^2, T)}\,,\ee
where $G_P(T)$ is the coupling constant appropriate for our study
of $\pi$ mesons. We have found $G_P(T)=13.49$\gev{-2} by fitting
the pion mass in a calculation made at $T=0$, with $m_u = m_d
=0.364$ GeV. The result given in Eq.\,(A18) is only expected to be
useful for small $P^2$, since the Gaussian regulator strongly
modifies the large $P^2$ behavior. Therefore, we suggest that the
following form is useful, if we are to consider the larger values
of $P^2$. \be \frac{C_{P}(P^2,
T)}{P^2}=\left[\frac{\tilde{J}_P(P^2, T)}{P^2}\right]
\frac{1}{1-G_P(T)J_P(P^2, T)}\,.\ee (As usual, we put
$\vec{P}=0$.) This form has two important features. At large
$P_0^2$, ${\mbox{Im}\,C_{P}(P_0, T)}/{P_0^2}$ is a constant, since
${\mbox{Im}\,\tilde{J}_{P}(P_0^2, T)}$ is proportional to $P_0^2$.
Further, the denominator of Eq.\,(A19) goes to 1 for large
$P_0^2$. On the other hand, at small $P_0^2$, the denominator is
capable of describing resonant enhancement of the correlation
function. As we will see, the results obtained when Eq.\,(A19) is
used appear quite satisfactory. (\,We may again refer to
Ref.\,[\,16\,], in which a similar approximation is described.)

For a study of the vector-isovector correlators, we introduce conserved vector currents $j_{\mu,
i}(x)=\tilde{q}(x)\gamma_{\mu}\lambda_i q(x)$ with i=1, 2 and 3. In this case we define \be
J_V^{\mu\nu}(P^2, T)=\left(g\,{}^{\mu\nu}-\frac{P\,{}^\mu P\,{}^\nu}{P^2}\right)J_V(P^2, T)\ee and
\be C_V^{\mu\nu}(P^2, T)=\left(g\,{}^{\mu\nu}-\frac{P\,{}^\mu P\,{}^\nu}{P^2}\right)C_V(P^2,
T)\,,\ee taking into account the fact that the current $j_{\mu,\,i}(x)$ is conserved. We may then
use the fact that \be J_V(P^2,T) = \frac13g_{\mu\nu}J_V^{\mu\nu}(P^2,T)\ee and
\be\mbox{Im}\,J_V(P^2,T)&=& \frac23\left[\frac{P_0^2+2m_u^2(T)}{4\pi}\right]
\left(1-\frac{4m_u^2(T)}{P_0^2}\right)^{1/2}e^{-\vec
k\,{}^2/\alpha^2}[\,1-2n_1(k)\,]\\
&\simeq& \frac{2}{3}\mbox{Im}J_P(P^2,T)\,.\ee (See Eq.\,(A7) for
the specification of $k=|\vec k|$.) We then have \be
C_V(P^2,T)=\tilde{J}_V(P^2,T)\frac1{1-G_V(T)J_V(P^2,T)}\,,\ee
where we have introduced \be\mbox{Im}\tilde{J}_V(P^2,T)&=&
\frac23\left[\frac{P_0^2+2m_u^2(T)}{4\pi}\right]
\left(1-\frac{4m_u^2(T)}{P_0^2}\right)^{1/2}[\,1-2n_1(k)\,]\\
&\simeq& \frac{2}{3}\mbox{Im}\tilde{J}_P(P^2,T)\,. \ee In the literature, $\omega$ is used instead
of $P_0$ [\,\,1-3\,\,]. We may define the spectral functions \be\sigma_V(\omega,
T)=\frac{1}{\pi}\,\mbox{Im}\,C_V(\omega, T)\,,\ee and \be\sigma_P(\omega,
T)=\frac{1}{\pi}\,\mbox{Im}\,C_P(\omega, T)\,,\ee

Since different conventions are used in the literature [\,1-3\,],
we may use the notation $\overline{\sigma}_P(\omega, T)$ and
$\overline{\sigma}_V(\omega, T)$ for the spectral functions given
there. We have the following relations: \be
\overline{\sigma}_P(\omega, T)=\sigma_P(\omega, T)\,,\ee and
\be\frac{\overline{\sigma}_V(\omega,
T)}{2}=\frac{3}{4}\sigma_V(\omega, T)\,,\ee where the factor 3/4
arises because, in Refs. [\,1-3\,], there is a division by 4,
while we have divided by 3, as in Eq.\,(A22).





\end{document}